\documentclass[journal]{IEEEtran}

\usepackage{graphicx} 
\usepackage{amsmath}  
\usepackage{hyperref} 
\usepackage{comment}
\usepackage{placeins} 

\begin{document}

\title{Developing Cost-Effective Drones for 5G Non-Terrestrial Network Research and Experimentation}
\author{Carlos de Quinto C\'{a}ceres, Andr\'{e}s Navarro, Alejandro Leonardo Garc\'{i}a Navarro, Tom\'{a}s Mart\'{i}nez, \\Gabriel Otero, Jos\'{e} Alberto Hern\'{a}ndez\\ Universidad Carlos III de Madrid, Spain\\ email: \{cquinto, annavarr, agnavarr\}@pa.uc3m.es, tmcortes@ing.uc3m.es, \{gaoterop, jahgutie\}@it.uc3m.es}
\date{June 2024}




\maketitle

\begin{abstract}
In this article, we describe the components and procedures for building a drone ready for networking experimentation. In particular, our drone design includes multiple technologies and elements such as 4G/5G connectivity for real-time data transmission, a 360-degree camera for immersive vision and AR/VR, precise GPS for navigation, and a powerful Linux-based system with GPU for computer vision experiments and applications. Component selection and assembly techniques are included, along with software integration for a smooth, seamless operation of advanced edge applications. 
\end{abstract}


%
\IEEEpeerreviewmaketitle

\section{Introduction}

\IEEEPARstart{T}{his} article provides a comprehensive guide for building (in a Do-It-Yourself fashion) a small drone for testing and evaluating connectivity in the context of Non-Terrestrial Networks (NTNs) for 5G and beyond, similar to other works found in literature~\cite{tal2021drone,mesquita2021steps,ginsberg2021auto,typeset2024materials,vayuyaan2024diy}. NTNs refer to communication systems that operate beyond the Earth's surface, mainly using satellites, high-altitude platform stations (HAPS) like stratospheric balloons, airships, Low-Altitude Platforms (LAPs) like drones or unmanned aerial vehicles (UAVs)~\cite{3gpp2024ntn,li2024unmanned}. 

NTNs have various applications such as providing broadband Internet in underserved areas, enabling maritime and aviation communications, supporting emergency response and disaster relief efforts, facilitating Internet of Things (IoT) connectivity, and serving as backhaul for terrestrial networks~\cite{mesquita2021steps,rajabifard2021potential,mary2022applications}. 

For example, about 20\% of US population lives in rural areas, which account for about 97\% of the total land~\cite{DELPORTILLO2019123}. This number grows to 28\% in Europe, and about 40\% world-wide. In many cases, fiber deployment does not reach rural areas (at least the last mile), since this results very expensive for network operators, hard to justify in terms of Average Revenue per User (ARPU). Indeed, it is estimated that every single meter of fiber connectivity costs approximately \$100 USD. The largest
share of this cost includes digging, trenching and the civil works
in general~\cite{schneir_2014}. NTNs can help provide broadband connectivity to such isolated areas where fiber deployment is not feasible~\cite{nwaogu2023application,mary2022applications,drones7080515}.

Concerning satellites, another important piece of the NTN ecosystem, the research community has witnessed a race toward deploying different satellite constellations to provide connectivity globally, with more than 50,000 satellites estimated to be launched within 10 years~\cite{mckensey_2020}. Thanks to the cost reduction in launching Low-Earth Orbit (LEO) satellites~\cite{techno_economic_sats}, four major companies are already deploying LEO satellite mega-constellations, namely Telesat, Tesla's Starlink, OneWeb and Amazon Kuiper~\cite{DELPORTILLO2019123}. However, there are a number of challenges related with the integration of satellites into the 5G ecosystem, mainly due to latency and doppler effects, as analyzed in~\cite{guidotti_LTE,guidotti_5g,aranity_2021}. A detailed survey on this matter is exhaustively studied in~\cite{ieeeaccess_survey}. Hence, satellites are expected to be also complemented with LAPS and UAVs like drones, which are inexpensive, easy to deploy and operate, and can be landed for operations maintenance and take off as quickly as needed and as many times as necessary. 

In the upcoming sections, we provide a short guide for building a medium-sized drone for 5G connectivity experiments and use cases. This includes the component selection and assembly procedures for building a drone, along with open-source software libraries needed to have it up and running. 

This drone is designed to weigh 6.5~Kg and cost less than 4,000 USD (at the time of writing, sept 2024), having an autonomy of 40 minutes. Such features open up a wide array of applications across various fields:

\begin{itemize}
    \item Advanced Surveillance and Reconnaissance: The combination of 5G connectivity, 360-degree camera, and GPS allows for real-time, high-resolution surveillance with a panoramic view. This makes it ideal for law enforcement, search and rescue operations, or monitoring large-scale events.
    \item AI-Powered Environmental Monitoring: Utilizing the Jetson Orin's processing power and onboard AI models, the drone can perform real-time analysis of environmental data. It could be used to monitor wildlife populations, track deforestation, or assess the impact of natural disasters.
    \item Smart Agriculture: The drone's capabilities make it an excellent tool for precision agriculture and Internet of Things (IoT) applications. It can analyze crop health, detect pests or diseases, and even assist in targeted application of fertilizers or pesticides.
    \item Autonomous Inspection of Infrastructure: Thanks to computer vision and image processing state-of-the-art software, drones can now perform detailed inspections of bridges, power lines, wind turbines, or other large structures, identifying potential issues without human intervention.
    \item Enhanced Film making and Photography: The 360-degree camera and stable flight characteristics make this drone an exceptional tool for cinematographers and photographers, offering unique perspectives and immersive footage.
    \item Edge Computing for Scientific Research: The onboard GPU and ability to run large language models allow for complex data processing in the field, providing valuable data for scientific expeditions in remote areas, enabling real-time analysis of collected data.
    \item Emergency Response and Disaster Management: The drone's 5G connectivity and advanced imaging can provide critical real-time information to first responders and disaster management teams, helping to coordinate relief efforts more effectively.
    \item Interactive Art Installations: The drone's combination of visual capabilities and processing power opens up possibilities for creating dynamic, AI-driven art installations or performances.
\end{itemize}

These applications demonstrate how the integration of advanced hardware and AI capabilities can transform a drone from a simple flying camera into a versatile, intelligent platform capable of tackling complex tasks across numerous industries.


The remainder of this work is organized as follows: Section \ref{hardware} explains the necessary hardware components required for building such a drone; and Section \ref{assembling} details the step-by-step process of assembling these components. Section~\ref{sec:extensions} provides some ideas for adding extra features to the drone, on attempts to build 5G NTN use cases. Finally, Section~\ref{sec:summary} concludes this work with a summary of its main contributions and take-away messages.

\section{Hardware design}
\label{hardware}

\subsection{Initial design and requirements}

This guide begins by emphasizing the importance of clarifying the primary objective of the drone. Whether intended for filming purposes or prioritizes endurance, this initial consideration serves as the cornerstone for developing a professional-grade UAV. The following list outlines the functional requirements the drone should fulfill, such as emergency fire detection, Large-Language Model support, or video recording. Other special technical specifications and requirements before starting the design process are:  

\begin{itemize}
    \item Aerodynamics and propulsion efficiency
    \item Weight distribution and balance
    \item Flight stability and control
    \item Power management and battery life
    \item Payload capacity
    \item Environmental durability
\end{itemize}

Our drone is designed to have about 45 minutes of autonomy, weigh 6.5 kilograms, and require 600 Watts of power.

\begin{figure*}
    \centering
    \includegraphics[width=0.7\textwidth]{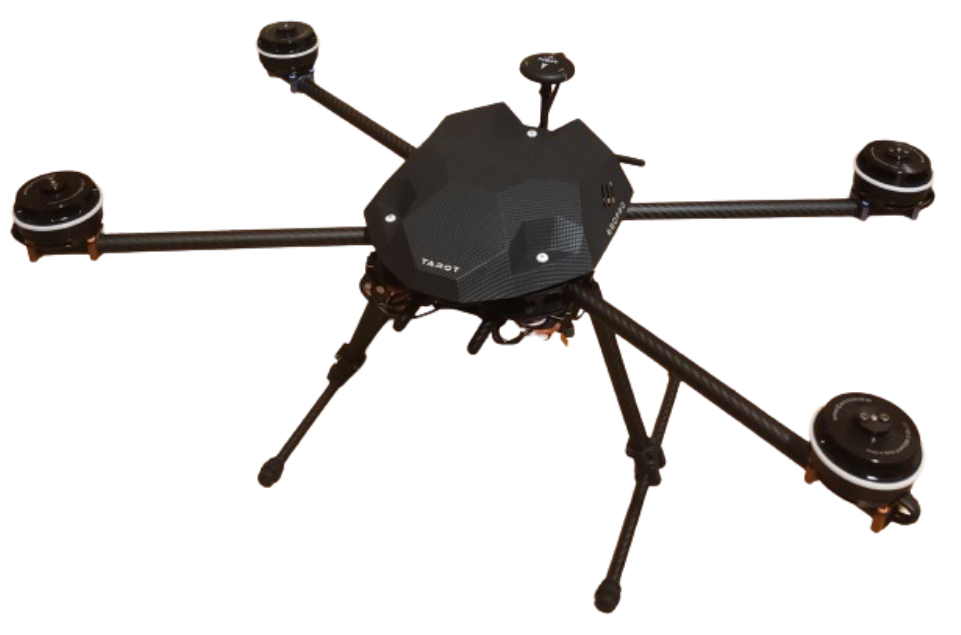}
    \includegraphics[width=\columnwidth]{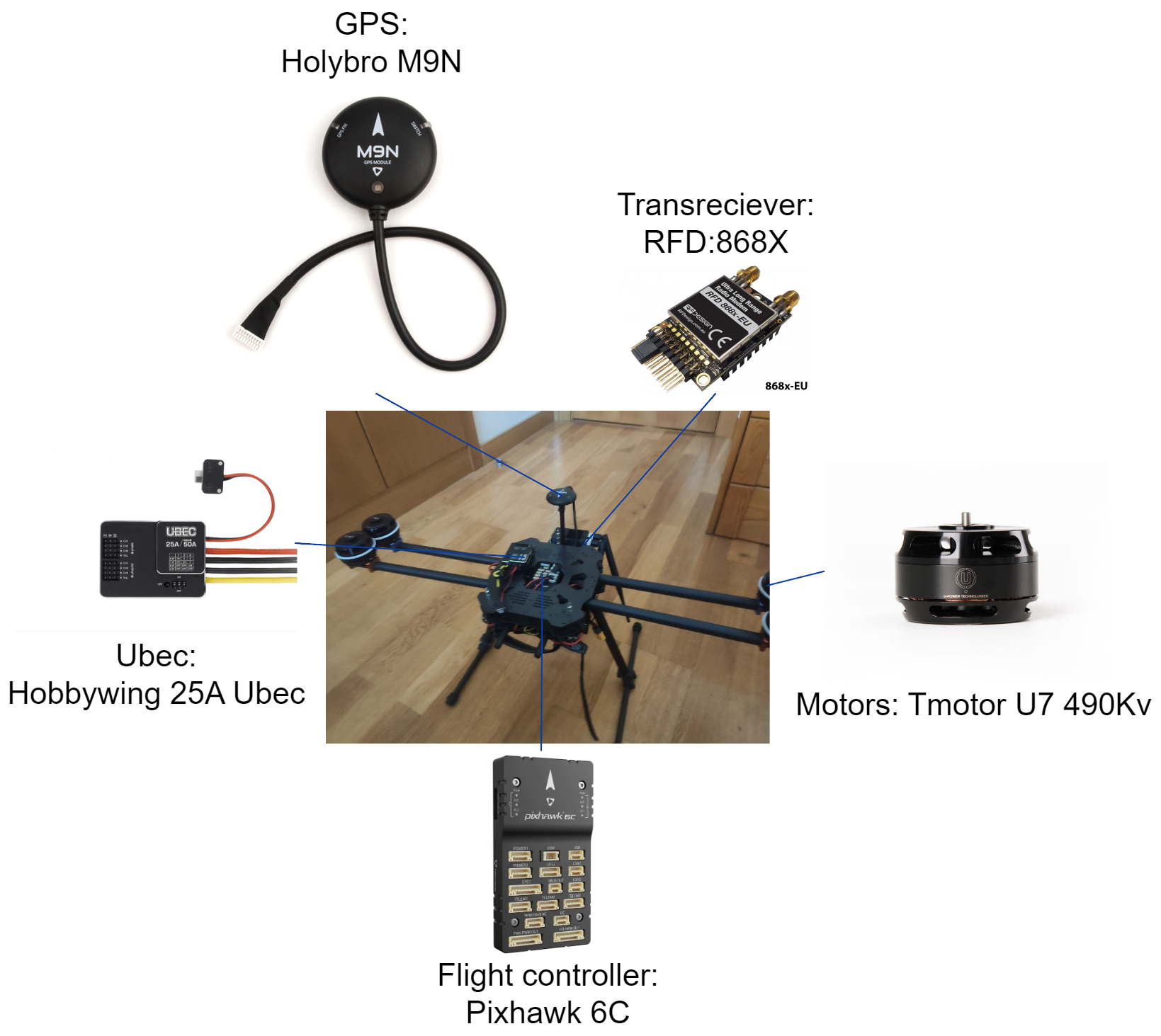}
    \includegraphics[width=\columnwidth]{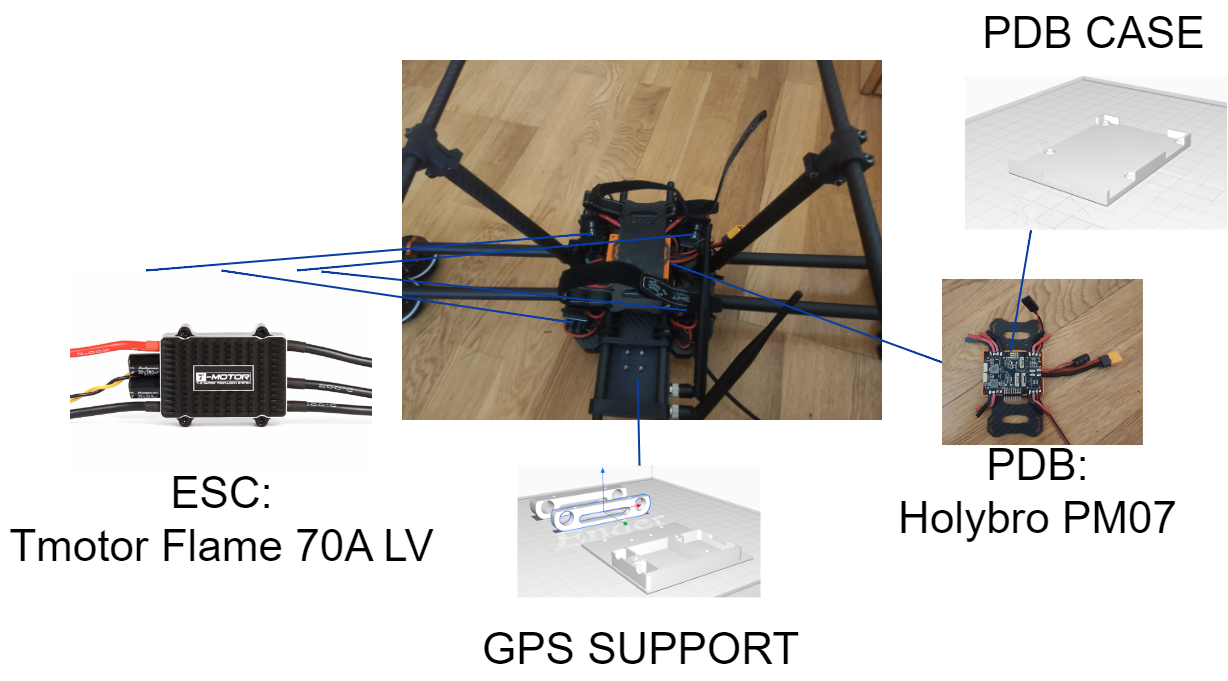}
    \caption{Drone finally assembled (top); components and top view (bottom left), and components and bottom view (bottom right).}
    \label{fig:dronefinal}
\end{figure*}

Fig.~\ref{fig:dronefinal} shows the final assembled drone along with the individual components on top and bottom. 

Also, Table~\ref{tab:components} provides a summary list of individual components, and the next section briefly overviews each hardware component in detail.

\begin{table*}[tbp]
    \centering
    \begin{tabular}{|c|cccc|}
    \hline
    Component & Description & Specifications & Approximate cost & Link amazon \\
    \hline
    
    Quadcopter Platform & Tarot XS690 & 17inch propeller &  200\$ & \href{https://a.co/d/cc5RtsL}{Link} \\

    Motors &  T-motor U7 V2 490KV &  & 150\$x4 &\href{https://a.co/d/8CD9ywW}{Link} \\
    
    ESC & T-motor Flame 70A LV ESC & 4-6S voltage, 70A continuous & 70\$ &\href{https://a.co/d/8Z4mZfy}{Link} \\
    
    FC, GPS and PDB combo & Pixhawk 6C with PDB and GPS & GPS Precision 1.5m CEP & 500\$ &\href{https://a.co/d/drJLg8Z}{Link}\\

    Propellers & Tmotor 17x5.8 V2 & 17 Inch carbon propellers & 72\$x2 &\href{https://rc-innovations.es/en/shop/tmotor-17x5-8-v2-pair-tmotor-17x5-8-v2-t-motor#attr=967,543,4905,3787}{Link}\\
    
    Telemetry and RC Control & RFD868 TXMOD V2 868Mhz & 1W 40km of range & 412\$ &\href{https://a.co/d/eyc937a}{Link} \href{https://a.co/d/4hxzaZx}{Link2} \\

    BEC & HobbyWing Ubec 25A HV & 25A  5V, 6V, 7.4V or 8.4V & 57\$ &\href{HobbyWing Ubec 25A HV}{Link}\\

    Charger & ISDT DUAL K4 & 400W & 230\$ &\href{https://a.co/d/7qgRDfQ}{Link}\\

    Battery & TATTU 22000mAh 14.8V 30C &  Li-po,EC5, 1677g  & 300\$ &\href{https://www.gensace.de/tattu-22000mah-14-8v-30c-4s1p-lipo-battery-pack-with-ec5-2212.html}{Link}\\

    Radiomaster & TX16S MAX ELRS & ELRS, EdgeTX & 220\$ &\href{https://a.co/d/e8QdZp3}{Link}\\

    Radiomaster battery & Radiomaster 2s 5000mah  & Li-Ion, Xt30, 7.4V & 30\$ &\href{https://a.co/d/gQobpAa}{Link}\\
    
    Lidar proximity sensor (optional) & MakerFocus Lidar Range Finder & 0,2 to 8M, unidirectional & 25\$ &\href{https://a.co/d/g0Zfzkv}{Link}\\      
    \hline
    \end{tabular}
        \caption{Individual components, approximate cost (as of Sept 2024) and links}
    \label{tab:components}
\end{table*}

Before starting, it is important to have:
\begin{itemize}
    \item Welding equipment
    \item A 3D printer
\end{itemize}

Next, we outline the basic building components used in the drone of Fig.~\ref{fig:dronefinal}, namely frame and body, motors and propellers, flight controller, battery, sensors (e.g. GPS, accelerometers) and camera/payload systems, among others.



\subsection{Platform}
Selecting the appropriate platform is a pivotal decision in the drone-building process, with the frame being the initial consideration. Given the goal of constructing a long-endurance drone capable of carrying substantial weight, we decided to opt for a quadcopter configuration. This choice is driven by the fact that quadcopters can accommodate larger propellers compared to hexacopters, thereby enhancing overall efficiency.

While hexacopters excel in speed and stability, quadcopters offer superior endurance and efficiency, aligning more closely with the desired characteristics for this particular project's functional requirements.


For this, we have chosen the 17-inch propeller Tarot XS690.

\subsection{Motors}

Selecting the right motors is crucial for our quadcopter's performance, and is nearly as important as selecting the frame. After extensive research and evaluation, we have chosen to implement motors from T-motor, an industry leader renowned for its high-quality multirotor propulsion systems.

From T-motor's lineup, the T-motor U7 V2 490KV has emerged as our optimal choice, since it offers an exceptional balance of power, efficiency, and versatility. Key features include:
\begin{enumerate}
    \item Versatile power handling: Capable of efficiently driving 17-inch propellers using either 4S (14.8V) or 6S (22.2V) LiPo batteries.
    \item Adaptive performance: The 490KV rating allows for a wide operational range, balancing thrust output with power consumption.
    \item Robust construction: Featuring high-quality materials and precise manufacturing for durability and reliability.
    \item Thermal efficiency: Advanced design for optimal heat dissipation, ensuring consistent performance during extended flight times.
\end{enumerate}

The U7 V2 490KV's adaptability is particularly advantageous for our project, as it allows us to fine-tune our power system based on specific mission requirements. We can optimize for either extended flight time using 4S batteries or maximize thrust with 6S configurations, without needing to change motors.



\subsection{Electronic Speed Controller}

Now that we have selected our motors, the next critical component is choosing a suitable Electronic Speed Controller (ESC). The ESC is vital as it regulates power delivery from the battery to the motors, ensuring smooth and precise control. Continuing with our T-motor ecosystem, we have selected the T-motor Flame 70A LV ESC.

Key features of the T-motor Flame 70A LV ESC include:

\begin{itemize}
    \item Current Rating: 70A continuous current, allowing for high-power handling capability.
    \item Low Voltage (LV) Compatibility: Specifically designed for use with lower voltage setups, including 4S LiPo batteries (14.8V nominal).
    \item Voltage Range: Supports 3S to 6S LiPo batteries (11.1V to 25.2V), providing flexibility in power system design.
    \item Heat dissipation: Optimized for heat dissipation and durability.
\end{itemize}

It is worth noting that we have selected the LV (Low Voltage) variant of the Flame ESC series. This choice is deliberate and essential for our build, as we intend to use 4S (14.8V) LiPo batteries. The LV models are optimized for these lower voltage ranges, ensuring efficient operation and proper motor control.

In contrast, the HV (High Voltage) models in the Flame series are designed for higher voltage systems, typically supporting 6S (22.2V) batteries and above. Using an HV ESC with our 4S setup would not work as there is not enough voltage supplied to power the components.

By pairing the T-motor U7 V2 490KV motors with the Flame 70A LV ESCs, we create a well-matched and efficient power system. This ESC selection complements our motor choice, resulting in a reliable and versatile power system for our quadcopter drone.





\subsection{Battery}

We intend to incorporate a Jetson Orin as the primary processing unit, weighing 750~g, alongside a router weighing 600~g, resulting in a combined weight of 1,350~g. Due to the high energy consumption of these components, it is crucial to include a substantial battery to offset power demands and sustain flight duration.

Considering the motors, ESC, frame, and other components, the total weight of the drone is 6.4~kg. Ideally, the drone should hover at half its maximum power. Motors operating on 4S configuration generate 1.1~kg of thrust at 50\% throttle, necessitating a 900~g battery for the entire system.

A 900~g battery is insufficient for this UAV application. Therefore, adjustments must be made to accommodate a larger battery. Options include reducing the drone's weight or enhancing its power output (utilizing 6S configuration). However, to maintain build simplicity, we opt to operate at 60\% throttle with a larger battery, albeit sacrificing some flight performance. Given the intended use of the drone for slow movements and hover predominantly, this compromise is deemed acceptable.

To address this, we will utilize a 4S 22Ah battery, weighing 2.5~kg, resulting in a total weight of 6.5kg. Consequently, each motor will be required to pull 1.5~kg during hover.

\subsection{Power Distribution Board}



For our high-performance drone, we have chosen the Holybro PM07 Power Distribution Board (PDB). This PDB is ideal for our current 4S setup and future-proofed for potential 6S upgrades. Notable features are:

\begin{enumerate}
    \item Voltage support up to 14S LiPo
    \item High current handling: 90A continuous, 140A peak
    \item Integrated voltage and current sensing (up to 140A)
    \item Built-in LC filter for clean power output
    \item Multiple output options:
    \begin{itemize}
        \item 8 pairs of ESC solder pads
        \item 5V 3A FC output
    \end{itemize}
    \item Compact 68*50*10mm design
\end{enumerate}

The PM07 meets our requirements for battery sensing without a separate sensor. This PDB ensures our drone's power system is robust, flexible, and capable of handling both current needs and future upgrades, making it an excellent fit for our build.

\subsection{Power for additional electronics}

Powering the electronic components of our drone, particularly the Jetson Orin and router, requires careful consideration to ensure stable and clean power delivery. 

For the Jetson Orin, our most valuable and power-hungry component, we have selected the Hobbywing UBEC 25A. This choice may seem excessive, as the Jetson Orin typically requires around 10V and 3A. However, our decision is rooted in future-proofing and expandability. The Hobbywing UBEC 25A offers:

\begin{enumerate}
    \item High current capacity: 25A continuous output
    \item Adjustable voltage: 5V/6V/7.4V/8.4V/9V/10V
    \item Wide input voltage range: 7V to 26V
    \item Built-in safety features: Over-current, over-temperature, and short-circuit protection
\end{enumerate}

This robust UBEC ensures clean, stable power to the Jetson Orin, crucial for its optimal performance and longevity. It also provides headroom for potential future upgrades or additional components.

For the router, which has different voltage requirements, we have opted for a standard BEC (Battery Elimination Circuit). This more modest BEC is sufficient for the router's power needs, balancing cost-effectiveness with reliable performance.

By using separate power regulation systems for these key components, we ensure each receives the appropriate, clean power supply, enhancing the overall reliability and performance of our drone's electronic systems.


\subsection{Flight Controller}

The flight controller is the central nervous system of our drone, responsible for processing sensor data, executing flight algorithms, and managing overall drone behavior. After careful consideration of several options including the Cube Orange, CUAV V5, and others, we have selected the Pixhawk 6C for our build.

Key features of the Pixhawk 6C include:
\begin{enumerate}
\item Powerful STM32H7 processor (480 MHz, 2 MB Flash, 1 MB RAM)
\item Integrated vibration isolation
\item Multiple connectivity options (USB-C, CAN, UART, I2C, SPI)
\item Dedicated safety switch and buzzer ports
\item Analog battery sensing input compatible with our PM07 PDB
\end{enumerate}

The Pixhawk 6C offers an excellent balance of performance and cost-effectiveness. Its analog battery sensing input is particularly valuable, allowing seamless integration with our chosen PM07 Power Distribution Board.

For the flight control firmware, we have chosen ArduPilot~\cite{ardupilot2024github}. This open-source platform supports various drone configurations, multiple sensors, and peripherals and is fully compatible with the most popular ground control stations.

ArduPilot's ongoing development and broad capabilities make it an ideal choice for our project, providing a robust and flexible foundation for our drone's flight control system. Its open-source nature also allows for customization if needed in future iterations of our build.


\subsection{Transmitter and Receiver}

For our drone's remote control system, we have selected the Radiomaster TX16S Max, widely recognized as the industry standard in RC controllers. This choice ensures we have a reliable, feature-rich interface for piloting our drone. The TX16S Max boasts a high-resolution color display for clear telemetry data and precise hall effect gimbals for accurate control inputs. Its customizable switches and buttons allow for versatile flight mode selection, while the EdgeTX firmware offers extensive customization options. The controller's multi-protocol support enables compatibility with various receivers, and an SD card slot facilitates easy firmware updates and model storage.

To enhance the capabilities of our control system, we have paired the TX16S Max with the TXMOD V2 module, incorporating the RFD868X system. This combination significantly extends our control range by operating on the 868MHz band, offering superior long-range performance compared to traditional 2.4GHz systems. The integrated telemetry feature provides real-time flight data transmission back to the controller, including crucial information like battery voltage, GPS coordinates, and altitude.

The RFD868X system employs frequency-hopping spread spectrum (FHSS) technology, ensuring a stable connection even in noisy RF environments. This robust link is crucial for maintaining control and data integrity during flight. Furthermore, the bidirectional communication capability allows for in-flight parameter adjustments and mission planning updates, adding a layer of flexibility to our drone operations.

By combining the Radiomaster TX16S Max with the TXMOD V2 and RFD868X, we have created a comprehensive control and telemetry solution in a single package. This setup offers the reliability, range, and flexibility required for our advanced drone application, ensuring we can maintain precise control and receive crucial flight data even in challenging conditions or at extended distances.


\subsection{Final additional components}

Lastly, before starting the assembly, a few additional parts are needed:

\begin{itemize}
    \item 3.5mm bullet connector
    \item 4mm heatshrink
    \item 20 AWG wire
    \item 14 AWG wire
    \item JR style connectors
    \item JST GH 6 pin and 5 pin connectors
    \item 10 AWG wire (optional)
\end{itemize}

\section{A guide for assembling the drone}
\label{assembling}

\subsection{Step 1: The PDB}

\begin{figure}
    \centering
    \includegraphics[width=\columnwidth]{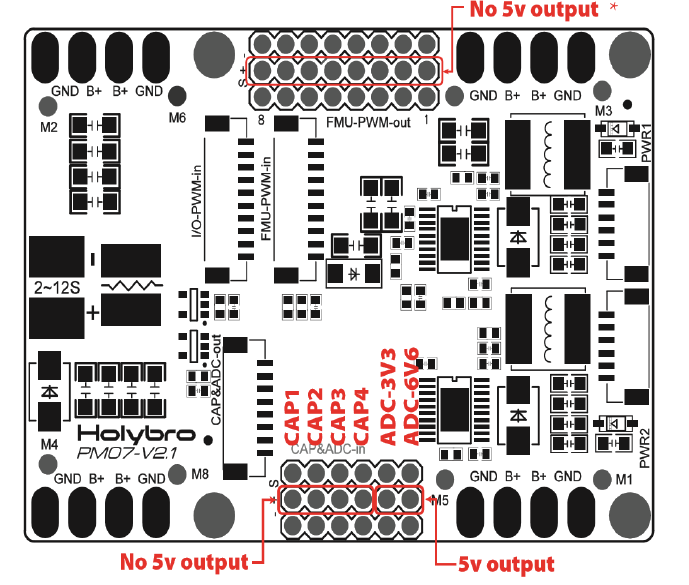}
    \includegraphics[width=\columnwidth]{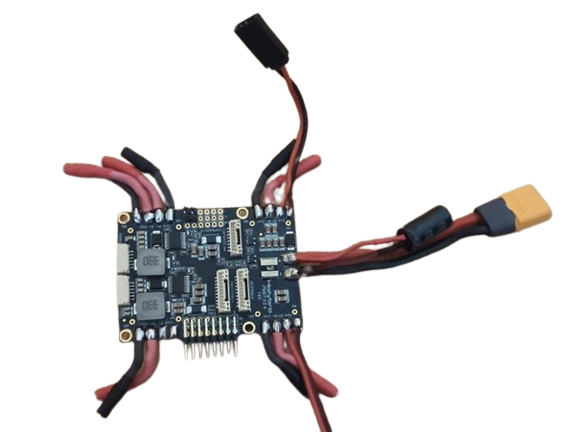}
    \caption{Step 1. PDB Diagram (top); PDB with soldered connectors (bottom).}
    \label{fig:step1}
\end{figure}

The first step to start with is the PDB: PM07 Power module (see Fig.~\ref{fig:step1}, left). Our drone’s power distribution system is designed with both current needs and future expandability in mind. The heart of this system is the Power Distribution Board (PDB), which will manage seven distinct battery outputs.

Four of these outputs are dedicated to the Electronic Speed Controllers (ESCs), which regulate power to our drone’s motors. These connections are critical for flight performance and require robust wiring. We are using 14 AWG (American Wire Gauge) wire for these connections, which offers low resistance and can handle the high current draw of our motors. Each wire will be terminated with a 3.5mm female bullet connector, allowing for secure yet easily detachable connections to the ESCs.

Another 14 AWG wire will power the Hobbywing BEC (Battery Elimination Circuit), which provides regulated power to our Jetson Orin. This thick gauge ensures minimal voltage drop, which is crucial for the stable operation of our main computing unit.

For our retractable landing gear and an additional expandability port, we are using 20 AWG wire. This gauge is sufficient for these lower-current applications. These wires will be fitted with JR-style connectors, a standard in RC applications, ensuring compatibility with a wide range of components.

The physical layout of these connections on the PDB is crucial for weight distribution and ease of maintenance. The four ESC wires will be soldered to the B+ (positive) and GND (ground) pads at each corner of the PDB. The placement of the BEC wire should be chosen based on the physical layout of components in the drone, minimizing wire length where possible.

Lastly, we're addressing a potential frame clearance issue by relocating the PDB's capacitor. We will carefully remove it and extend its leads, allowing us to reposition it without compromising its crucial function of smoothing voltage fluctuations.

This comprehensive power distribution setup provides a solid foundation for our drone, ensuring reliable power delivery to all components while maintaining flexibility for future modifications or additions. By the end of this process, the PDB setup should resemble Fig.~\ref{fig:step1}, right.

It is important to note that you might consider replacing these bullet connectors for XT60s if you are concerned about plugging peripherals in with the wrong polarity. However, since we don’t have a lot of space, we have opted for this option, being very careful about the polarity of the components.

\subsection{Step 2: The frame}

\begin{figure}
    \centering
    \includegraphics[width=0.7\columnwidth]{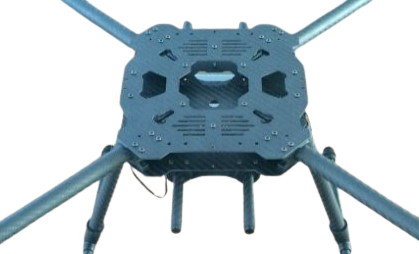}
    \includegraphics[width=0.5\columnwidth]{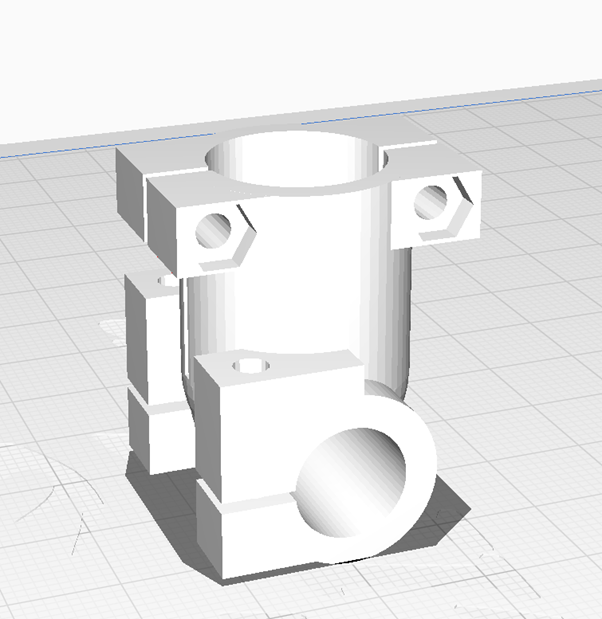}
    \includegraphics[width=0.8\columnwidth]{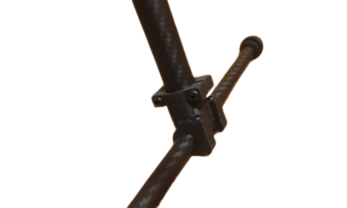}
    \caption{Step 2. Tarot frame (top); Landing Gear in 3D (middle), and printed part landing gear (bottom).}
    \label{fig:step2}
\end{figure}

Transitioning to the frame and motors, our initial task is to assemble the frame (Fig.~\ref{fig:step2}, left) according to the provided image, omitting the assembly of the arms or the top plate. Although no specific instructions are provided, the assembly process is straightforward.

Before assembling the frame, we need a 3D printer for certain parts. Specifically, the landing gear T-junction on this frame is notably flimsy, prompting the creation of a more durable and reliable replacement requiring 4 M3 bolts and nuts for assembly (see Fig.~\ref{fig:step2}, middle and right).

\subsection{Step 3: Motors}

\begin{figure}
    \centering
    \includegraphics[width=0.81\columnwidth]{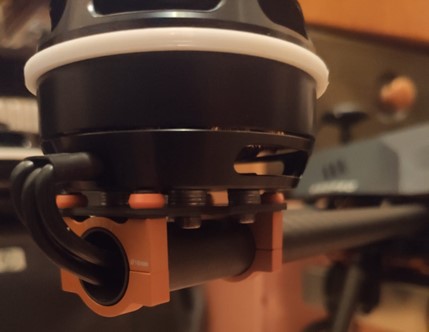}
    \includegraphics[width=0.81\columnwidth]{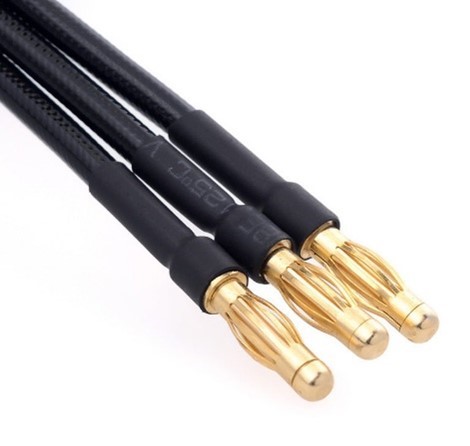}
    \caption{Step 3. Detail of the motor mounting solution (top); Motor Connectors (bottom).}
    \label{fig:step3}
\end{figure}

Moving forward to the motors, we need to affix each motor onto its motor mount and subsequently onto the arms. Given the size of these motors and the motor mounts, we must  incorporate four washers with each screw to ensure clearance of the motor mount bolts, as depicted in Fig.~\ref{fig:step3} (left).

Next, we weld 3.5mm male bullet connectors to the ends of the cables, as demonstrated in Fig.~\ref{fig:step3} (right). These cables are then folded inside the arms and drawn out from the other end.

With these preparations completed, we can now proceed to mount the arms onto the frame using the bottom screws. Special care needs to be taken as the frame is now very fragile, since it does not have the strength of the top plate.

\subsection{Step 4: Preparation of the ESC}

Next we need to prepare the ESCs to be mounted onto the frame. Since these ESCs already come with female bullet connectors welded to the outputs, we only need to weld bullet connectors to the + and – cables of the ESCs and the Hobbywing BEC (or XT60s for added safety):

\begin{figure}[h!]
    \centering
    \includegraphics[width=0.5\columnwidth]{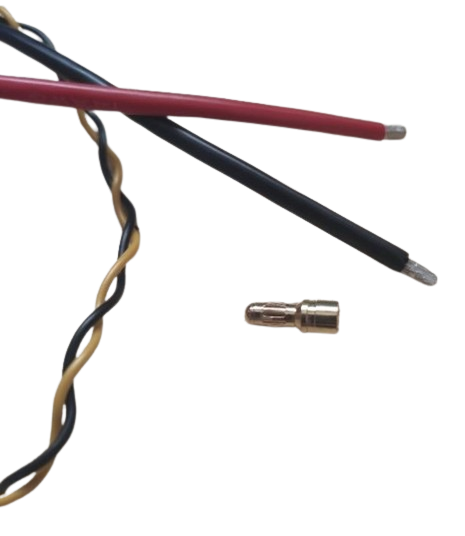}
    \caption{Step 4. ESC connectors.}
    \label{fig:ESC}
\end{figure}

\subsection{Step 5. Other cables}
\begin{figure}
    \centering
    \includegraphics[width=0.8\columnwidth]{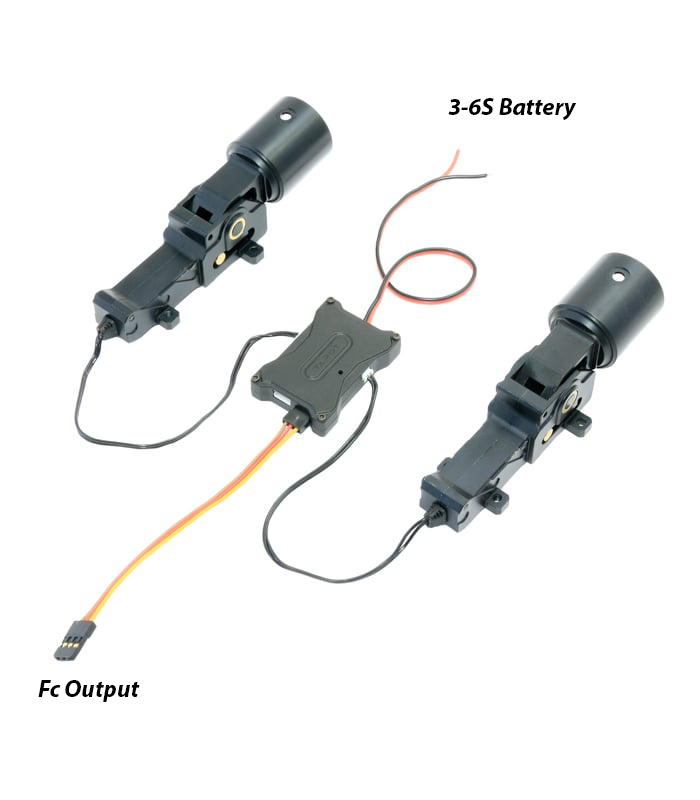}
    \includegraphics[width=0.8\columnwidth]{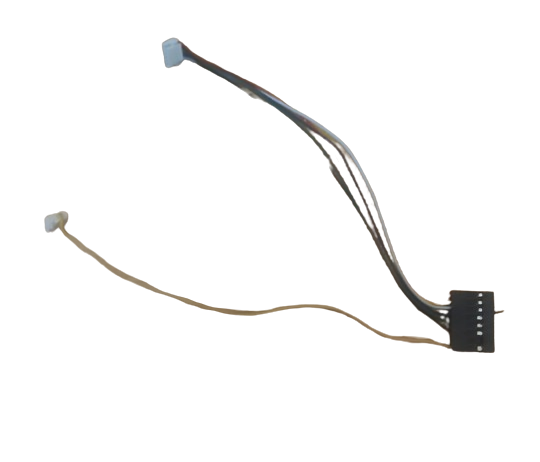}
    \caption{Step 5. Landing gear electronics (top); RFD Cable (bottom).}
    \label{fig:cables}
\end{figure}

Now we need to make a cable that can connect to the power port of the PDB that we previously created (see Fig.~\ref{fig:cables}, left). We also need a servo cable to connect the signal of the FC to the box, as shown on the left.

Additionally, we need to make a cable to establish the connection between the Pixhawk 6C and the RFD receiver, such as the one in Fig.~\ref{fig:cables}, right.

\subsection{Step 6. Step-by-step Component Placement}

\begin{figure*}[h!]
    \centering
    \includegraphics[width=0.7\columnwidth,angle=90]{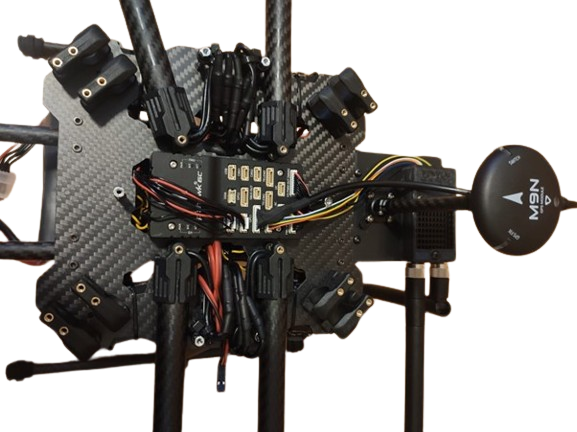}
    \includegraphics[width=0.5\columnwidth]{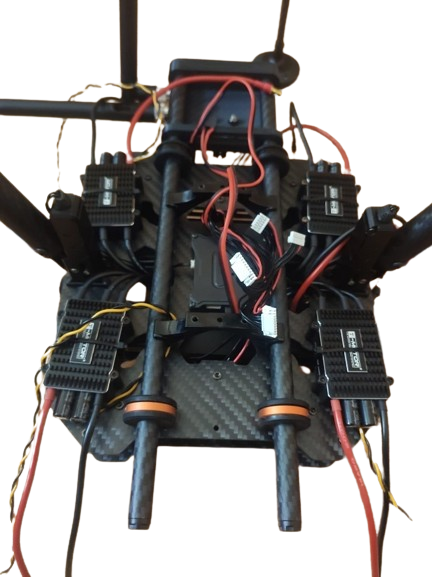}
    \caption{Flight controller centered on the frame (perspective 1); Flight controller centered on the frame (perspective 2).}
    \label{fig:placement}
\end{figure*}

The selected approach optimizes frame utilization through strategic component placement. The flight controller (FC) is centrally mounted as illustrated in Fig.~\ref{fig:placement} (perspective 1), with full component integration pending.

FC attachment utilizes four Velcro squares, providing vibration dampening and positional stability. Precise central positioning is critical, avoiding arm contact.

Careful cable management is implemented for the FMU port, I/O PWM out, and power cables, routing them beneath the FC.

The drone is inverted for ESC installation on the bottom plate's underside and landing gear electronic box mounting.

ESC securement employs a combination of zip ties and double-sided adhesive. Zip tie diameter is optimized to maintain arm rotational freedom. The resultant configuration is depicted in Fig.~\ref{fig:placement} (perspective 2).

\begin{figure*}
    \centering
    \includegraphics[width=0.9\columnwidth]{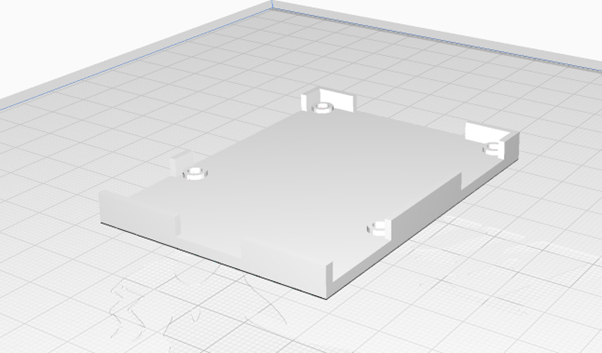}
    \includegraphics[width=0.7\columnwidth]{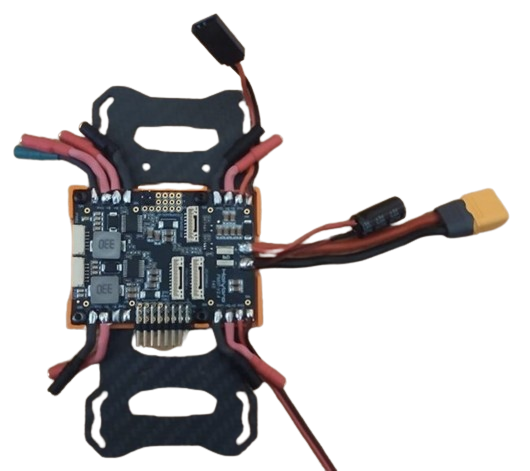}
    \caption{The PDB case (left); PDB Mounted in the case (right).}
    \label{fig:placement2}
\end{figure*}

PDB installation follows, positioned between the bottom and battery plates. A custom 3D-printed mount, as shown in Fig.~\ref{fig:placement2} (left), provides electrical isolation. High-strength double-sided adhesive secures the case to the battery plate (Fig.~\ref{fig:placement2}, right).

\begin{figure*}
    \centering
    \includegraphics[width=0.5\columnwidth]{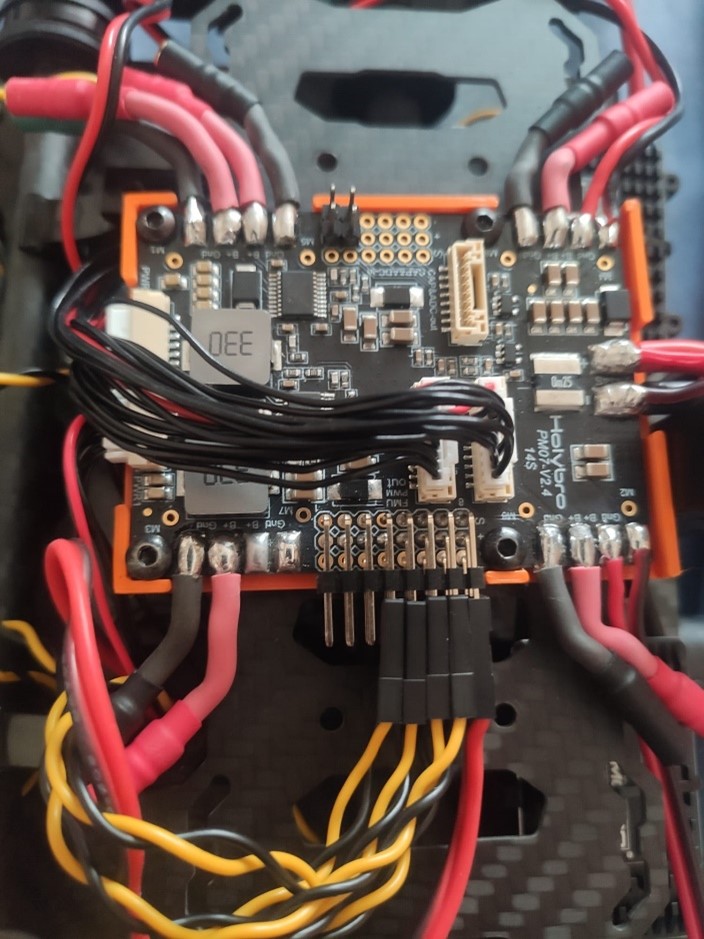}
    \includegraphics[width=0.7\columnwidth,angle=90]{Fotos/Flight_controller_Centered__sin_background_.png}
    \caption{PDB with all the connections (left); Flight controller centered on the frame (right).}
    \label{fig:placement3}
\end{figure*}

ESC power cables (black and yellow) are connected to the PDB's FMU/OUT port and the landing gear power cable. ESC placement is finalized, ensuring power cable accessibility. The configuration should mirror Fig.~\ref{fig:placement3} (left).

The plate is affixed to the frame's provided plastic components, with the PDB-containing section oriented inward.

Post plate securement (via four M3 bolts), the ESC and Hobbywing BEC are connected with strict polarity observation. Meticulous cable management is crucial to prevent pinching or damage.

ESC outputs are connected to motors, with careful wire routing to avoid arm interference during frame folding. The final configuration is illustrated in Fig.~\ref{fig:placement3} (right).

Wire positioning is confined between the metal standoffs limiting arm movement.

\begin{figure*}
    \centering
    \includegraphics[width=\columnwidth]{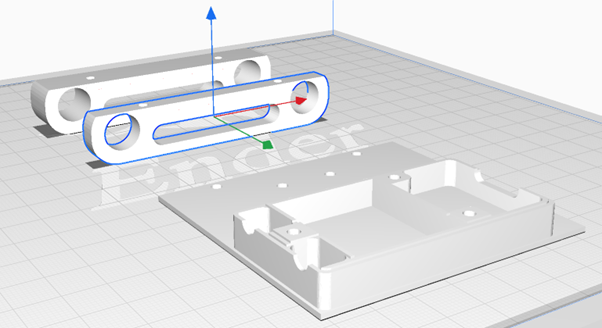}
    \includegraphics[width=0.4\columnwidth]{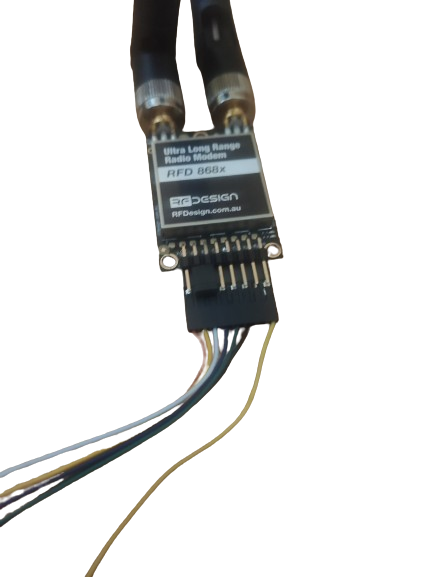}
    \caption{STL files for the GPS and telemetry module mount (left); RFD module connected (right).}
    \label{fig:placement4}
\end{figure*}

RFD868X and GPS integration follows, utilizing a custom 3D-printed mount (Fig.~\ref{fig:placement4}, left) for optimal spacing.

The pre-fabricated cable is connected to the module prior to case installation, as shown in Fig.~\ref{fig:placement4} (right).

Module power is sourced from the FC, with reduced power settings to mitigate potential FC damage. While a separate 5V BEC would be optimal for full-power operation, the current configuration suffices at 20 dBm for the intended application.

\begin{figure*}
    \centering
    \includegraphics[width=0.8\columnwidth]{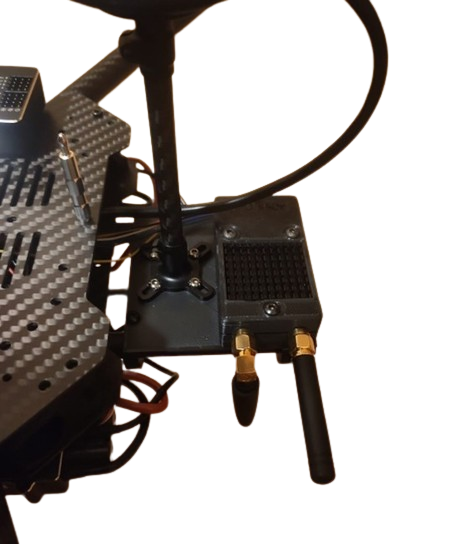}
    \includegraphics[width=0.8\columnwidth]{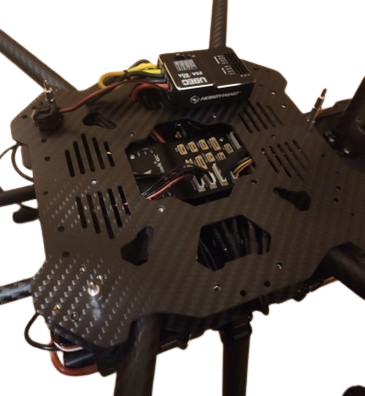}
    \caption{GPS and RFD module connected (left); UBEC placement and top plate (right).}
    \label{fig:placement5}
\end{figure*}

Final module placement and connection is executed (Fig.~\ref{fig:placement5}, left).

Top plate installation follows, with consideration for BEC placement and cable routing for PDB Bat+ and Bat- connections (Fig.~\ref{fig:placement5}, right).

Securement of top screws completes the assembly process, ensuring component stability and overall structural integrity.

\section{Extensions for 5G NTN research and experiments}
\label{sec:extensions}

\begin{figure}
\centering
\includegraphics[width=0.65\columnwidth]{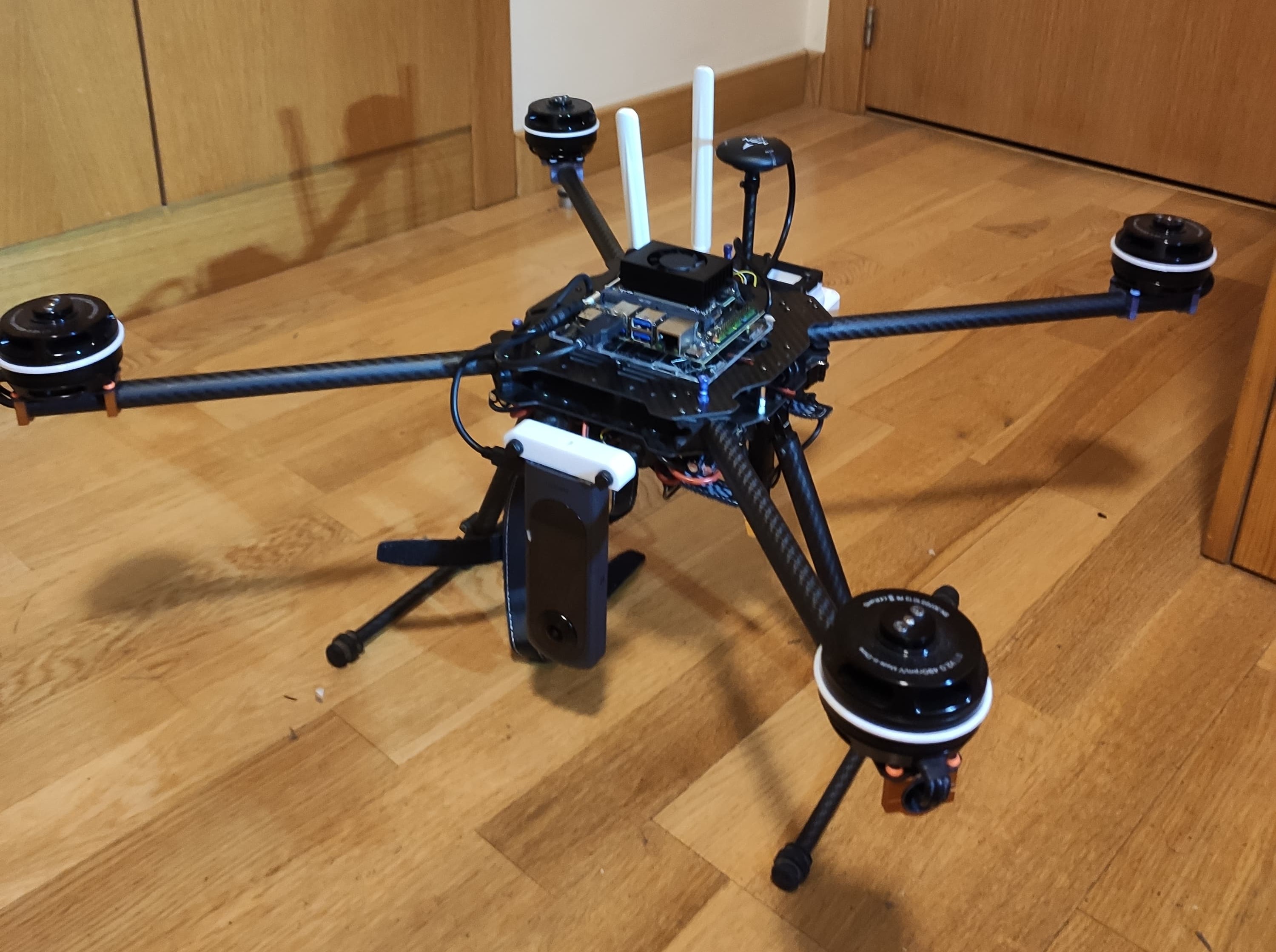}
\includegraphics[width=0.65\columnwidth]{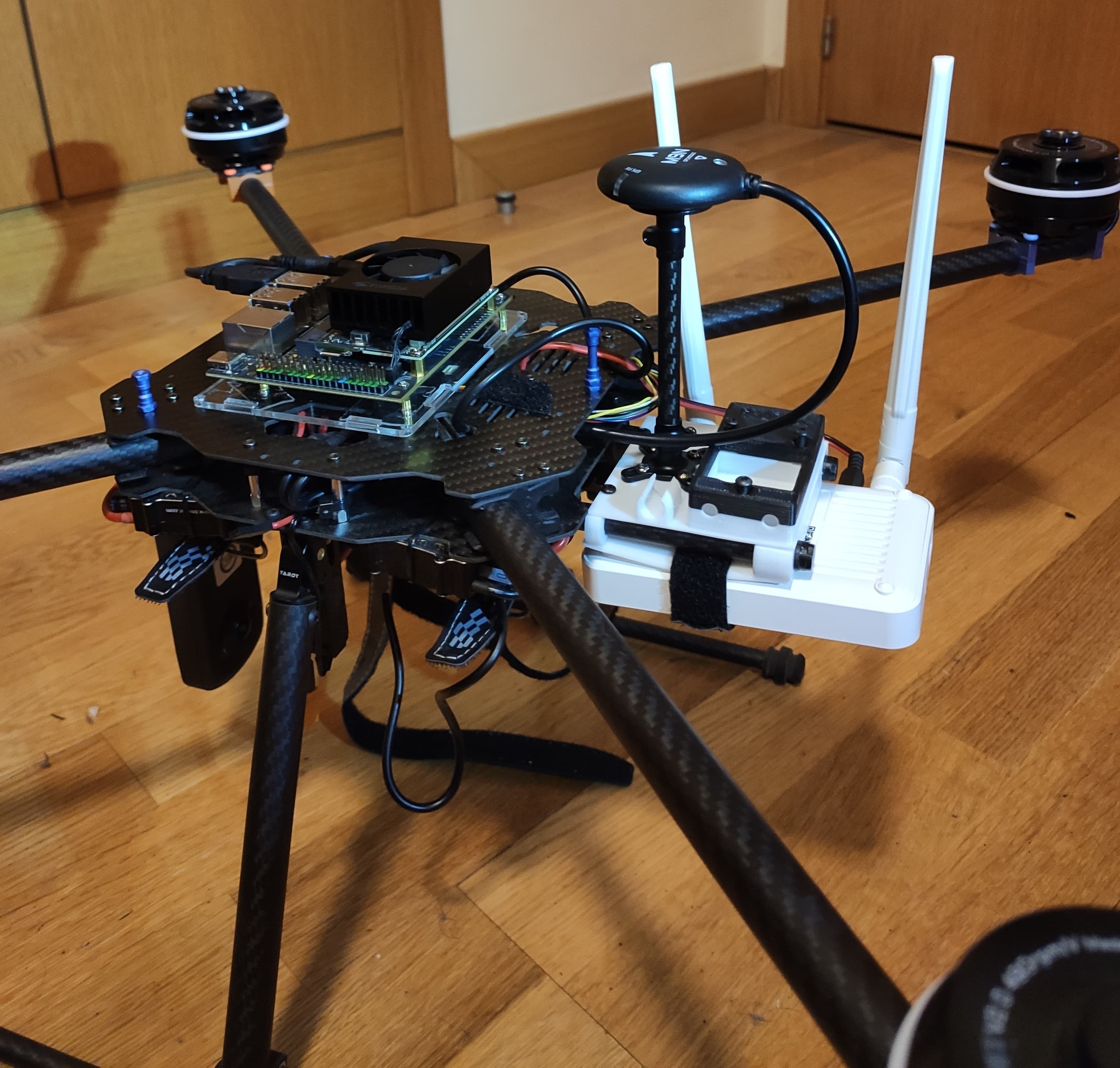}
\caption{Front and rear views of 5G component integration}
\label{fig:Componentes 5g}
\end{figure}

Component integration necessitated modifications to several 3D-printed parts to accommodate the electronic modules. The 360-degree camera is secured via a standardized 1/4"-20 UNC tripod mount. The Jetson compute module is affixed using a custom-designed base, adhered to the carbon fiber frame with high-strength double-sided adhesive tape due to the absence of compatible mounting points. The router is secured using a tension-adjustable battery strap, with its position strategically placed between the antennas to prevent lateral displacement, as illustrated in Fig.~\ref{fig:Componentes 5g}.

To facilitate high-bandwidth communication between the UAV and ground control station (GCS), a 5G router has been integrated into the system. This solution provides the requisite data throughput for real-time telemetry and payload data transmission. While latency is a consideration, it is not critical for this application, and we anticipate slightly elevated latency values during operation.

A key objective of this project is to establish a low-latency, high-fidelity video link between an onboard 360-degree camera and the pilot. The Ricoh Theta X was selected for its ability to output full-resolution imagery via USB interface. The video pipeline involves streaming the camera feed to the onboard NVIDIA Jetson Orin compute module, where an RTSP (Real-Time Streaming Protocol) server is configured to facilitate connection between the drone and pilot.

The Jetson Orin, running a customized Linux distribution, leverages its GPU capabilities for edge computing and computer vision tasks. This setup enables on-board processing of visual data, potentially including real-time fire detection algorithms implemented using Python libraries optimized for NVIDIA CUDA cores.

A summary of this is provided in Table~\ref{tab:components_extra}.

\begin{table*}[htbp]
    \centering
    \begin{tabular}{|c|cccc|}
    \hline
    Component & Description & Specifications & Approximate cost & Link amazon \\
    \hline 
    
    Nvidia Jetson Orin & Onboard computer for image detection & 8GB RAM  5.3 FP32 TFLOPs & 500\$  &\href{https://a.co/d/6DhUbEc}{Link} \\

    Ricoh Theta X & 360 streaming camera & 5.7K, 30fps  & 800\$ & \href{https://a.co/d/e1urhuR}{Link}\\

     (GL-X750V2) Smart WiFi & 4G/5G router & Dual band, OpenWrt & 150\$ & \href{https://a.co/d/fmMgUxM}{Link}\\
    \hline
    \end{tabular}
        \caption{Extra components for 5G NTN research and experiments.}
    \label{tab:components_extra}
\end{table*}


\section{Summary and discussion}
\label{sec:summary}

This article provides a comprehensive guide for building a cost-effective drone specifically designed for 5G Non-Terrestrial Network (NTN) research and experimentation. We detail the process of constructing a drone that integrates multiple technologies including 4G/5G connectivity, a 360-degree camera, precise GPS, and a powerful Linux-based system with GPU capabilities. 

In a nutshell, the article includes: 
\begin{itemize}
    \item Detailed hardware specifications and component selection, including the frame, motors, electronic speed controllers, and other essential parts.
    \item Step-by-step assembly instructions for building the drone.
    \item Integration of advanced features such as 5G connectivity, 360-degree camera, and edge computing capabilities.
    \item Potential applications of the drone in various fields, including surveillance, environmental monitoring, smart agriculture, and emergency response.
\end{itemize}

The resulting drone is cost-effective since it has been designed to weigh 6.5 kg, and has an autonomy of 40 minutes. With an estimated cost of under $4,000$ USD, this drone represents a significant cost saving compared to many commercial alternatives, potentially enabling more widespread adoption in research settings. Its versatile deign, including features like 5G connectivity, 360-degree camera, and edge computing capabilities, opens up possibilities for its use across various disciplines. The DIY approach allows researchers to tailor the drone to their specific needs, which could be particularly valuable in the rapidly evolving field of 5G NTN research. Finally, this guide could also serve as an excellent educational resource for students and early-career researchers in telecommunications, robotics, and related fields.



\section*{Acknowledgments}
The authors would like to acknowledge the support of Spanish projects ITACA (PDC2022-133888-I00) and 6G-INTEGRATION-3 (TSI-063000-2021-127).


\bibliographystyle{IEEEtran}  
\bibliography{satellites,PONs,drones}  

\begin{thebibliography}{10}
\providecommand{\url}[1]{#1}
\csname url@samestyle\endcsname
\providecommand{\newblock}{\relax}
\providecommand{\bibinfo}[2]{#2}
\providecommand{\BIBentrySTDinterwordspacing}{\spaceskip=0pt\relax}
\providecommand{\BIBentryALTinterwordstretchfactor}{4}
\providecommand{\BIBentryALTinterwordspacing}{\spaceskip=\fontdimen2\font plus
\BIBentryALTinterwordstretchfactor\fontdimen3\font minus \fontdimen4\font\relax}
\providecommand{\BIBforeignlanguage}[2]{{%
\expandafter\ifx\csname l@#1\endcsname\relax
\typeout{** WARNING: IEEEtran.bst: No hyphenation pattern has been}%
\typeout{** loaded for the language `#1'. Using the pattern for}%
\typeout{** the default language instead.}%
\else
\language=\csname l@#1\endcsname
\fi
#2}}
\providecommand{\BIBdecl}{\relax}
\BIBdecl

\bibitem{tal2021drone}
D.~Tal and J.~Altschuld, \emph{Drone Technology in Architecture, Engineering and Construction: A Strategic Guide to Unmanned Aerial Vehicle Operation and Implementation}.\hskip 1em plus 0.5em minus 0.4em\relax Wiley, 2021.

\bibitem{mesquita2021steps}
G.~P. Mesquita, J.~D. Rodr{\'\i}guez-Teijeiro, and R.~R. de~Oliveira, ``Steps to build a diy low-cost fixed-wing drone for biodiversity conservation,'' \emph{PLoS ONE}, vol.~16, no.~8, p. e0255559, 2021.

\bibitem{ginsberg2021auto}
\BIBentryALTinterwordspacing
S.~Ginsberg, ``Auto drone: Modifying a diy drone kit for autonomous flight,'' Washington University in St. Louis, Independent Study, 2021. [Online]. Available: \url{https://openscholarship.wustl.edu/cgi/viewcontent.cgi?article=1143&context=mems500}
\BIBentrySTDinterwordspacing

\bibitem{typeset2024materials}
Typeset.io, ``What materials and tools are necessary for building a diy drone?'' \url{https://typeset.io/questions/what-materials-and-tools-are-necessary-for-building-a-diy-193xmrt9hy}, 2024, accessed: September 27, 2024.

\bibitem{vayuyaan2024diy}
Vayuyaan, ``Diy drone - exciting guide to building your drone,'' \url{https://vayuyaan.com/blog/diy-drone-exciting-guide-to-building-your-drone/}, 2024, accessed: September 27, 2024.

\bibitem{3gpp2024ntn}
\BIBentryALTinterwordspacing
{3GPP}, ``Non-terrestrial networks (ntn),'' 3rd Generation Partnership Project, Tech. Rep., 2024, accessed: September 27, 2024. [Online]. Available: \url{https://www.3gpp.org/technologies/ntn-overview}
\BIBentrySTDinterwordspacing

\bibitem{li2024unmanned}
Y.~Li, J.~Ren, N.~Cheng, W.~Shi, and X.~Shen, ``Unmanned autonomous intelligent system in 6g non-terrestrial networks: A survey,'' \emph{Information}, vol.~15, no.~1, p.~38, 2024.

\bibitem{rajabifard2021potential}
A.~Rajabifard, G.~Foliente, and D.~Paez, ``The potential of drone technology in pandemics,'' in \emph{COVID-19 Pandemic, Geospatial Information, and Community Resilience}.\hskip 1em plus 0.5em minus 0.4em\relax Taylor \& Francis, 2021.

\bibitem{mary2022applications}
N.~Mary, S.~Shafiya, and M.~Ben~Maaouia, ``Applications of drone technology in construction projects: A systematic literature review,'' \emph{International Journal of Research - GRANTHAALAYAH}, vol.~10, no.~10, pp. 1--14, 2022.

\bibitem{DELPORTILLO2019123}
\BIBentryALTinterwordspacing
I.~{del Portillo}, B.~G. Cameron, and E.~F. Crawley, ``A technical comparison of three low earth orbit satellite constellation systems to provide global broadband,'' \emph{Acta Astronautica}, vol. 159, pp. 123--135, 2019. [Online]. Available: \url{https://www.sciencedirect.com/science/article/pii/S0094576518320368}
\BIBentrySTDinterwordspacing

\bibitem{schneir_2014}
J.~Rendon~Schneir and Y.~Xiong, ``Cost analysis of network sharing in ftth/pons,'' \emph{IEEE Communications Magazine}, vol.~52, no.~8, pp. 126--134, 2014.

\bibitem{nwaogu2023application}
J.~M. Nwaogu, Y.~Yang, A.~P. Chan, and H.-l. Chi, ``Application of drones in the architecture, engineering, and construction (aec) industry,'' \emph{Automation in Construction}, vol. 152, p. 104870, 2023.

\bibitem{drones7080515}
A.~not provided in the~search results], ``An overview of drone applications in the construction industry,'' \emph{Drones}, vol.~7, no.~8, p. 515, 2023.

\bibitem{mckensey_2020}
\BIBentryALTinterwordspacing
C.~Daehnick, I.~Klinghoffer, B.~Maritz, and B.~Wisem, ``Large leo satellite constellations: Will it be different this time?'' McKensey \& Company, Tech. Rep., May 2020. [Online]. Available: \url{https://www.mckinsey.com/industries/aerospace-and-defense/our-insights/large-leo-satellite-constellations-will-it-be-different-this-time}
\BIBentrySTDinterwordspacing

\bibitem{techno_economic_sats}
K.~T. Li, C.~A. Hofmann, H.~Reder, and A.~Knopp, ``A techno-economic assessment and tradespace exploration of low earth orbit mega-constellations,'' \emph{IEEE Communications Magazine}, pp. 1--7, 2022.

\bibitem{guidotti_LTE}
A.~Guidotti, A.~Vanelli-Coralli, M.~Caus, J.~Bas, G.~Colavolpe, T.~Foggi, S.~Cioni, A.~Modenini, and D.~Tarchi, ``Satellite-enabled lte systems in leo constellations,'' in \emph{2017 IEEE International Conference on Communications Workshops (ICC Workshops)}, 2017, pp. 876--881.

\bibitem{guidotti_5g}
A.~Guidotti, A.~Vanelli-Coralli, M.~Conti, S.~Andrenacci, S.~Chatzinotas, N.~Maturo, B.~Evans, A.~Awoseyila, A.~Ugolini, T.~Foggi, L.~Gaudio, N.~Alagha, and S.~Cioni, ``Architectures and key technical challenges for 5g systems incorporating satellites,'' \emph{IEEE Transactions on Vehicular Technology}, vol.~68, no.~3, pp. 2624--2639, 2019.

\bibitem{aranity_2021}
G.~Araniti, A.~Iera, S.~Pizzi, and F.~Rinaldi, ``Toward 6g non-terrestrial networks,'' \emph{IEEE Network}, vol.~36, no.~1, pp. 113--120, 2022.

\bibitem{ieeeaccess_survey}
F.~Rinaldi, H.-L. Maattanen, J.~Torsner, S.~Pizzi, S.~Andreev, A.~Iera, Y.~Koucheryavy, and G.~Araniti, ``Non-terrestrial networks in 5g \& beyond: A survey,'' \emph{IEEE Access}, vol.~8, pp. 165\,178--165\,200, 2020.

\bibitem{ardupilot2024github}
A.~Developers, ``Ardupilot: Arduplane, arducopter, ardurover, ardusub source code,'' \url{https://github.com/ArduPilot/ardupilot}, 2024, accessed: 2024-07-27.

\end{thebibliography}


\begin{thebibliography}{1}





\end{thebibliography}

%








\end{document}